# Enterprise Portal Development Tools: Problem-Oriented Approach

Sergey V. Zykov
ITERA Oil and Gas Company
Moscow, Russia
E-mail: szykov@itera.ru

**Abstract**[1]

The paper deals with problem-oriented visual information system (IS) engineering for enterprise Internet-based applications, which is a vital part of the whole development process. The suggested approach is based on semantic network theory and a novel *ConceptModeller* CASE tool.

## 1. Introduction: General Outlook of SDK

The present-day society has accumulated a huge and ever-growing data bulk, manipulating which is quite an issue, particularly, because of its heterogeneous and weak-structured character. Under socio-economic globalization, efficient information system (IS) functioning requires development of novel design and implementation concepts and methodologies. Thus, conceptually and methodologically comprehensive SDK should be created to integrate efficient IS maintenance throughout the entire lifecycle.

Research objective is analysis and development of conceptual and methodological foundations of full-scale IS development and its application to enterprise information collection, processing and reporting within global network environment. Major research tasks include generation of a conceptual approach for continuous integrated design, implementation and support for globally distributed portal-based IS by means of novel problem-oriented CASE-and-RAD-tools.

*ConceptModeller* CASE tool is aimed at semantically oriented visual IS development in a heterogeneous computational environment. The tool comprises components for frame notation [5] visualization, frame-to-UML conversion and for visualization of the resulting UML diagrams.

The order in which components have been named corresponds to the general IS development direction – from problem domain conceptual model entities to UML diagrams.

To provide heterogeneous (meta)database ((M)DB) unification and an easy control by *Microsoft .NET* platform, XML language has been chosen for (meta)data objects ((M)DO) definition. Therewith, the data structures provide convenient frame visualization and translation into UML language specification.

XML-based DB management is implemented using *XML Designer* component of *Microsoft Visual Studio 2005* SDK, which is used for DB template generation based on XML schema.

Let us overview *ConceptModeller* implementation by describing its components.

## 2. Visualization Component

The component is aimed at frame elements visualization and their (behavior) storage. Upon XML file opening/saving, element-by-element create/write operations for class instantiations is done by object-oriented functions and methods, written in C# programming language.

Double frame representation (both in graphical and structured DB form) requires frame storage DB format, which meets requirements for completeness, extendibility (for adding metadata) and visualization uniqueness (including file reopening operations).

Frame elements are characterized by the following attributes: ID, type, name, coordinates, predecessor and successor hierarchical links, and a number of optional fields.

Complete description of frame visualization DB scheme is stored in an XML file. Examples of internal and visual frame representations are given in fig.1 and fig.2 respectively.

The XML code fragment describes a frame instantiating `USER` concept by '`sergey.zykov`' value.

---

```xml
<?xml version="1.0"
         standalone="yes" ?>
- <NewDataSet>
- <Elements>
  <Id>1</Id>
  <Type>Var</Type>
  <Name>USER</Name>
  <Left>50</Left>
  <Top>70</Top>
  <Width>150</Width>
  <Height>80</Height>
  <Prev>0</Prev>
  <Next>0</Next>
  </Elements>
- <Elements>
  <Id>2</Id>
  <Type>Concept</Type>
  <Name>sergey.zykov</Name>
  <Left>50</Left>
  <Top>270</Top>
  <Width>150</Width>
  <Height>80</Height>
  <Prev>0</Prev>
  <Next>0</Next>
  </Elements>
- <Elements>
  <Id>4</Id>
  <Type>i</Type>
  <Name>i role</Name>
  <Left>125</Left>
  <Top>270</Top>
```

**Figure 1. XML frame description in visualization DB**

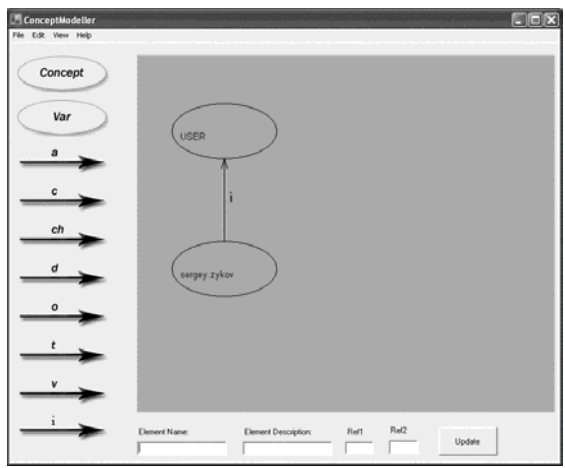

**Figure 2. Frame visual representation**

(M)DB scheme is stored in XML format.

For visibility sake, let us present a simplified (M)DB structure storage scheme for frame visualization (see fig.3).

```xml
<?xml version="1.0"
         encoding="utf-8" ?>
- <xs:schema id="NewDataSet"
      xmlns=""
      xmlns:xs="http://www.w3.org/
          2001/XMLSchema"
      xmlns:msdata="urn:schemas-
          microsoft-com:xml-msdata">
- <xs:annotation>
- <xs:appinfo source="urn:schemas-
   microsoft-com:xml-msdatasource">
- <DataSource
      DefaultConnectionIndex="0"
      Modifier="AutoLayout,
      AnsiClass, NotPublic, Public"
      xmlns="urn:schemas-microsoft-
      com:xml-msdatasource">
  <Connections />
  <Tables />
  <Sources />
  </DataSource>
  </xs:appinfo>
  </xs:annotation>
- <xs:element name="NewDataSet"
      msdata:IsDataSet="true"
      msdata:Locale="ru-RU">
- <xs:complexType>
- <xs:choice minOccurs="0"
      maxOccurs="unbounded">
- <xs:element name="Elements">
- <xs:complexType>
- <xs:sequence>
  <xs:element name="Id"
   type="xs:int" minOccurs="0" />
  <xs:element name="Type"
   type="xs:string"minOccurs="0" />
  <xs:element name="Name"
   type="xs:string"minOccurs="0" />
  <xs:element name="Left"
   type="xs:int" minOccurs="0" />
  <xs:element name="Top"
   type="xs:int" minOccurs="0" />
  <xs:element name="Width"
   type="xs:int" minOccurs="0" />
  <xs:element name="Height"
   type="xs:int" minOccurs="0" />
  <xs:element name="Prev"
   type="xs:int" minOccurs="0" />
  <xs:element name="Next"
   type="xs:int" minOccurs="0" />
  <xs:element name="Description"
   type="xs:string"minOccurs="0" />
  </xs:sequence>
  </xs:complexType>
  </xs:element>
  </xs:choice>
  </xs:complexType>
  </xs:element>
  </xs:schema>
```

**Figure 3. Generalized XML-scheme for frame visualization**

As one can see, each (M)DO is characterized by the following attributes: ID, type, name, description, visualization coordinates and pointers to preceding and succeeding objects.

Visualization module implementation size totals to around 0.4 Mbytes including nearly 0.25 Mbytes of basic (M)D schemes.

## 3. Frame-to-UML Translation Component

The component is aimed at automated translation of frame internal representation into UML notation. Frames describe situations in problem domains. Subsequent UML model processing by CASE tools during integrated software development is implied.

Let us note that the component allows bidirectional conversion (frame-to-UML and backwards), i.e. BPR software development is supported.

Total implementation time for the frame-to-UML translation module is 3 man-months.

*Microsoft .NET* has been chosen as implementation environment; *Microsoft Visual Studio 2005* has been used as supporting SDK.

The reasons for implementation environment choice are its Internet orientation and embedded efficient *XML Parser* component. C# language has been chosen for the implementation due to its efficiency and functional flexibility in the environment.

Component implementation size amounts to 500 Kbytes and around 2,500 source code lines.

The reason for such a large implementation size as compared to the previous component is bidirectional application development capability. The reason for short implementation term is comprehensive formal model features, which reduce BPR solution to reverse substitution.

Implementation benefits are based on the conceptual and methodological formal models and technologies underlying the integrated software development. Particularly, cross-platform internet-oriented IS development with language interoperability is used.

Therewith, labor expense savings for *ConceptModeller* and its components are reached thanks to environment-embedded parser and (M)DO scheme storage in XML and XDS formats, which can be easily visualized in *Windows and .NET* environments.

(M)DO scheme visualization example in *Microsoft Visual Studio 2005* environment is given in fig.4.

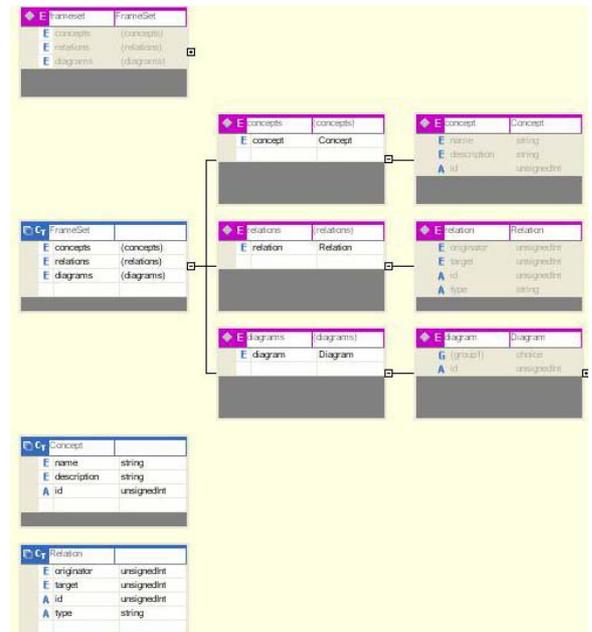

**Figure 4. (M)DB scheme fragment visualization**

As one can see from the above example, the visualization procedure represents an XML (M)DB fragment as a traditional UML class diagram interface, which contains atomic element descriptions (particularly, concepts and relationships represented by arcs) as well as larger objects (frame diagrams). Let us note that the description notation for the semantic network elements is fully object-oriented (e.g., a concept corresponds to a class containing such attributes as *ConceptModeller* internal ID, problem-oriented name and extended description).
Let us note that (M)DB schemes automated visualization by the suggested environment saves labor expenses of first-time IS development (at least by 25%), and it makes enterprise IS (re)engineering essentially easier (at least by 10%) for the users (chiefly system analytics) due to better unification, standardization and ergonomics.

## 4. Internal UML Format Visualization Component

*ConceptModeller* final operation stage for direct translation is (M)DO internal format visualization. The (M)DO have been produced by translation of frame-based family of problem domain (situation) models.

The component is primary oriented to UML class diagrams, the reason for this being frame translation mechanism features.

The computation environment *of Microsoft Visual Studio 2005* benefits automated UML visualization (due to *Microsoft Visio* integration, a tool aimed at general purpose business diagrams processing), which allows essential implementation terms reduction (by 50% and more). Therewith, high-level component-oriented *Microsoft Visio Tools for Visual Studio 2005* interfaces have been used. The implementation process involved Microsoft technical support Internet and e-mail hotlines.

The total component size is around 1,000 source code lines in C#.

Implementation term for UML diagram visualization component is two man-months. Since visualization is done by external *Microsoft Visio* tool in standard UML notation (version 2.0), the component screenshots are redundant.

## 5. Implementation Results Summary, Recommendations and Prospects

The paper summarizes problem-oriented IS development methodology for heterogeneous enterprise information resources.

To solve the task of enterprise information resources management, the full-scale *ConceptModeller* CASE tool has been designed and implemented on the basis of the author's conceptual and methodological approach and derived models, methods, architecture and interface solutions.

The CASE tool has been implemented in a globally distributed Internet and Intranet environment on the basis of UML and BPR IS development standards, Java and .NET technologies.

*ConceptModeller* has been implemented into enterprise integrated Internet IS development methodology. The implementation experience has proved shortening terms and reducing costs as compared to advanced commercial software available. It has also proved high adaptability, extendibility and ergonomics of the methodology.

Practical implementation experience proved actuality, innovation and efficiency of the approach on the whole and separate concepts, models and tools developed.

The author is going to continue his studies of the formal models that support enterprise portal-based IS problem-oriented development.